\documentclass[a4paper,11pt]{article}

\usepackage{pos}
\usepackage{lipsum} 

\usepackage{lineno}

\title{Measuring the Astrophysical $\bar{\nu}:\nu$ Ratio with IceCube}

\ShortTitle{Measuring the Astrophysical $\bar{\nu}:\nu$ Ratio with IceCube}

\author{The IceCube Collaboration* \\{\normalsize \normalfont(a complete list of authors can be found at the end of the proceedings)}\\}

\emailAdd{bskrzypek@lbl.gov}
\emailAdd{srklein@lbl.gov}

\abstract{

Recent measurements of astrophysical neutrinos have expanded our understanding of their nature and origin. However, very little is still known about the astrophysical $\bar{\nu}:\nu$
 ratio. The only prior measurement is the recent, single Glashow event seen by IceCube. Understanding the astrophysical $\bar{\nu}:\nu$
 ratio has a bearing on multiple questions, including the astrophysical spectral shape and neutrino production mechanisms. This analysis uses a new approach to measuring the astrophysical muon $\bar{\nu}:\nu$
ratio at various energies. It uses inelasticity, the fraction of the initial neutrino energy carried away by the hadronic shower. Inelasticity probes the $\bar{\nu}:\nu$
 ratio due to the fact that at energies below roughly 100 TeV, valence quarks dominate in deep inelastic scattering interactions, leading to different neutrino and antineutrino inelasticities and cross-sections. We use 10.3 years of IceCube data consisting of starting tracks at energies between 1 TeV and 1 PeV with a self-veto selection that enhances astrophysical event purity in the down-going direction. Based on this sample and analysis method, we present the first measurement of the astrophysical $\bar{\nu}:\nu$
 ratio at sub-PeV energies.

\vspace{4mm}

{\bfseries Corresponding authors:}
Barbara Skrzypek$^{1,2*}$, 
Spencer Klein$^{1,2}$\\
{$^{1}$ \itshape University of California, Berkeley}\\
{$^{2}$ \itshape Lawrence Berkeley National Laboratory}\\[4mm]
$^*$ Presenter
}

\ConferenceLogo{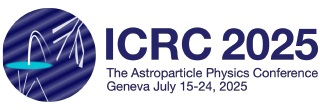}

\FullConference{39th International Cosmic Ray Conference (ICRC2025)\\
 15–24 July 2025\\
Geneva, Switzerland\\}

\begin{document}

\maketitle

\section{Introduction}

Despite multiple measurements of the high-energy astrophysical neutrino spectrum, the astrophysical $\bar{\nu}:\nu$ ratio remains largely unconstrained. This ratio is an important probe of neutrino production mechanisms in astrophysical sources and also affects how the astrophysical flux is inferred from measured neutrino interactions. 
For example, if neutrinos are produced in $pp$ interactions (where an accelerated proton collides with gas or dust), then we expect roughly equal numbers of $\pi^+$ and $\pi^-$ to be produced, leading to equal amounts of neutrinos and antineutrinos \cite{Barger:2014iua,IceCube:2021rpz}.  This mechanism is expected to be dominant in star-forming galaxies.  Because oscillations en-route to Earth mix the neutrino flavors, the $\bar{\nu}:\nu$ should be largely independent of flavor.

On the other hand, if neutrinos are formed when accelerated protons interact with photons ($p\gamma$ interactions), then proton excitation is an important reaction and many of the resulting neutrinos will come from the decay $\Delta^+\rightarrow \pi^+ n$. The resulting $\bar{\nu}:\nu$ for $\nu_e$, 3.5:1, is larger \cite{IceCube:2021rpz}. This includes $\nu$ from the pion decay and from the subsequent $\mu$ decay.  If the interaction occurs in a strong magnetic field, then the $\mu$ will lose energy and the subsequent $\bar{\nu}$ fraction will drop.  Conversely, if the interaction occurs in a region with a high accelerating gradient \cite{Klein:2012ug}, then the $\mu$ decay will dominate, and $\bar{\nu}:\nu = 1:1$.  Alternately, if the neutrinos are produced by neutron (or other nuclei) beta decay, then the flux will be nearly 100\% $\bar{\nu}$.


Few probes are sensitive to the astrophysical $\bar{\nu}:\nu$ ratio. One such example is the detection of an electron-antineutrino undergoing Glashow resonance \cite{IceCube:2021rpz}. This was detected by IceCube at 6 PeV with a significance of $5\sigma$. However, this observable only tests the electron-neutrino $\bar{\nu}:\nu$ component at very high energies. 

Inelasticity is another probe that offers sensitivity to other flavor components and over a wider energy range. In deep inelastic scattering (DIS) interactions, inelasticity is defined to be the fraction of the initial neutrino energy that gets carried away by the hadronic shower \cite{Klein:2019nbu}. At Cherenkov detectors, it is sensitive to charged current interactions with muons. Due to the fact that valence quarks, which dominate at lower energies, interact at different rates with neutrinos and antineutrinos, the corresponding cross-sections differ, leading to different inelasticities, particularly at lower energies. Previous IceCube inelasticity data were used to measure the atmospheric $\bar{\nu}:\nu$ ratio and also neutrino charm quark production \cite{IceCube:2018pgc}, and a recent IceCube measurement extended our understanding of inelasticity in neutrino DIS below $560\;{\rm GeV}$ \cite{IceCubeCollaborationSS:2025zgz}, finding good agreement with accelerator-based measurements. 
\section{Analysis}

In DIS, the charged-current cross-section for neutrinos and anti-neutrinos is given by \cite{Formaggio_2012}
\begin{equation}\label{eq:cross-section}
    \frac{d^2\sigma^{\nu,\bar{\nu}}}{dxdy} = \frac{G_F^2M_NE_{\nu}}{\pi(1+Q^2/M_W^2)^2}\Big[\frac{y^2}{2}2xF_1(x,Q^2)+\Big(1-y-\frac{M_Nxy}{2E_{\nu}}\Big)F_2(x,Q^2)\pm y\Big(1-\frac{y}{2}\Big)xF_3(x,Q^2)\Big],
\end{equation}
\noindent where the $+$($-$) sign corresponds to neutrino (anti-neutrino) interactions, $G_F = 1.17\times 10^{-5}\; {\rm GeV}^{-2}$, $M_N$ is the nucleon mass, and the kinematic variables are defined as 
\begin{eqnarray}
    x &=& Q^2/2M_NE_{\nu}y, \\
    y &=& E_{\rm had}/E_{\nu} = (E_{\nu}-E_{\ell})/E_{\nu},\\
    Q^2 &=& (p_{\nu}-k_{\ell})^2 = -m_{\ell}^2+2E_{\nu}(E_{\ell}-p_{\ell}\cos\theta_{\ell}).
\end{eqnarray}
\noindent The factors $F_i(x,Q^2)$ are the dimensionless nucleon structure functions. Under the quark-parton model, these can be expressed in terms of the corresponding quark parton distribution functions, which depend on the target and type of interaction involved. In the case of $\bar{\nu}p\rightarrow \mu^+X$ scattering, the \textit{leading order} structure functions can be expressed as a sum over the active quark parton distribution functions:
\begin{eqnarray}
    F_2^{W^-}(x,Q^2) &=& 2x(u+c+b+\bar{d}+\bar{s}+\bar{t}),\\
    xF_3^{W^-}(x,Q^2) &=& 2x(u+c+b-\bar{d}-\bar{s}-\bar{t}).
\end{eqnarray}
\noindent The structure functions corresponding to $\nu n\rightarrow \mu^-X$ scattering are obtained through the interchange $u \leftrightarrow d$. The third structure function, $xF_1(x,Q^2)$, is related to $F_2$ via a longitudinal structure function that quantifies the ratio of scattering off of longitudinally and transversely polarized $W$ bosons. In the quark-parton model at leading-order, this simplifies to $2xF_1 = F_2$ for both neutral- and charged-current interactions. 
Applying the leading order structure functions and ignoring terms proportional to $1/E_{\nu}$, we obtain the following differential cross-sections for neutrinos and anti-neutrinos: 
\begin{equation}
    \frac{d^2\sigma^{\bar{\nu}}}{dxdy} = \frac{G_F^2M_NE_{\nu}}{\pi(1+Q^2/M_W^2)^2}\Big[2x\Big(\bar{d}+\bar{s}+\bar{t}+(1-y)^2(u+c+b)\Big)\Big],
\end{equation}
\begin{equation}
    \frac{d^2\sigma^{\nu}}{dxdy} = \frac{G_F^2M_NE_{\nu}}{\pi(1+Q^2/M_W^2)^2}\Big[2x\Big(d+s+t+(1-y)^2(\bar{u}+\bar{c}+\bar{b})\Big)\Big].
\end{equation}
From the double-differential cross-section, we obtain inelasticity distributions by integrating over $y$ and normalizing by the total cross-section: 
\begin{equation}
    \frac{dp_{\nu}}{dy} = \frac{1}{\sigma_{\nu}}\int_0^1dx \frac{d^2\sigma_{\nu}}{dxdy},\quad \frac{dp_{\bar{\nu}}}{dy} = \frac{1}{\sigma_{\bar{\nu}}}\int_0^1dx \frac{d^2\sigma_{\bar{\nu}}}{dxdy}.
\end{equation}
To compute the probability distribution function corresponding to a scenario in which the ratio $\nu/(\nu+\bar{\nu}) = r$, we combine the individual probability distribution functions in the following way: 
\begin{equation}
    \frac{dp}{dy} = r\frac{dp_{\nu}}{dy}+(1-r)\frac{dp_{\bar{\nu}}}{dy}.
\end{equation}

\begin{figure}[h]
    \centering
    \includegraphics[width=0.49\linewidth]{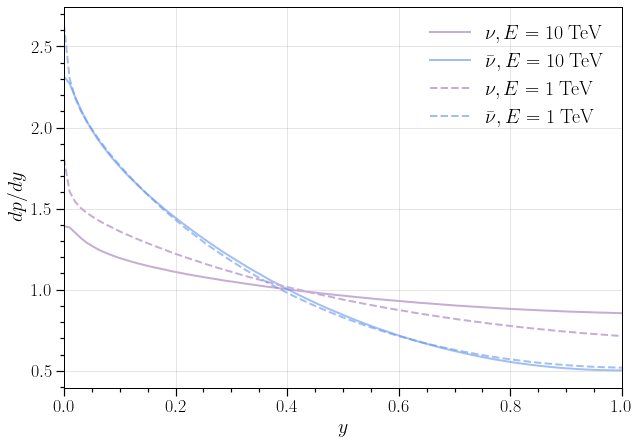}
    \includegraphics[width = 0.49\linewidth]{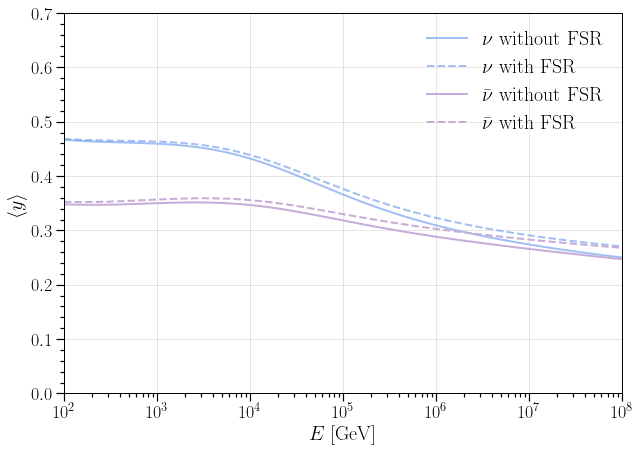}
    \caption{Left: The inelasticity probability distribution functions for neutrinos and antineutrinos at two different energies. Right: The mean inelasticity $\langle y \rangle \equiv \int dy y (dp/dy)$ as a function of energy. The dashed lines correspond to the mean inelasticity including a correction for final state radiation (FSR) \cite{FSR}.}
    \label{fig:dpdy}
\end{figure}

We use a recent set of leading-order quark parton distribution functions (PDFs). These are given by CTLO18A \cite{ct18} and documented in \texttt{LHAPDF} \cite{lhapdf}. At lower energies, the PDFs of the valence quarks dominate, leading to different quark and antiquark distributions. At higher energies, sea quarks begin to play a bigger role, and thus the difference between neutrino and antineutrino interactions becomes smaller.
The resulting inelasticity PDFs, defined as 
$ \frac {1}{\sigma }\frac {d\sigma }{dy}$, depend on the energy and the $\bar{\nu}:\nu$ ratio. An all-neutrino spectrum prefers larger inelasticities while an all-antineutrino spectrum exhibits smaller inelasticities.

To account for nuclear effects in DIS interactions in ice, we use nuclear PDFs for $^{16}{\rm O}$, given by EPPS21 and CT18A. We then combine the $\nu- ^{16}{\rm O}$ cross-section built out of these PDFs with the $\bar{\nu}-p$ cross-section, obtained using the nucleon PDFs described above, resulting in the cross-section for an ${\rm H}_2{\rm O}$ target.

\subsection{Data Sample}

The IceCube Neutrino Observatory is a cubic-kilometer Cherenkov detector located at the South Pole, $1.5$~km below the surface of the ice \cite{Halzen:2010yj,Aartsen_2017}. Strings running from below the ice to the computing lab consist of a total of 5160 digital optical modules (DOMs) that detect Cherenkov radiation from charged particles using photomultiplier tubes.

In order to measure the astrophysical $\nu/\bar{\nu}$ ratio, we use the IceCube Enhanced Starting Tracks Event Selection (ESTES) \cite{IceCube:2024fxo}. As a starting events sample, it offers comparatively good energy resolution and allows us to reconstruct inelasticity, while tracks provide good angular resolution.
Additionally, to remove the atmospheric muon background, ESTES makes use of a dynamic self-veto in the southern sky, which reduces background while retaining a significant number of astrophysical signal events.
We use the most recent 10.3 year sample with the dynamic self-veto. Our energies of interest range from 1 TeV to 10 PeV, and we probe neutrino directions covering the entire sky. We also use visible inelasticity \cite{IceCube:2018pgc}, which is defined as the fraction of the total energy reconstructed as a cascade.

To compute the direction of an event, a series of increasingly complex reconstruction algorithms are used, where the seed for each subsequent algorithm is the reconstructed direction of the previous algorithm. This includes \texttt{Millipede} \cite{Aartsen_2014}, which uses information from all available DOMs and divides tracks into segments, fitting an energy-loss for each segment so as to account for the stochastic energy loss of muons. 

We use an energy reconstruction based on a random forest algorithm \cite{random-forests}. This reconstructs the estimated initial hadronic and leptonic energies based on deposited track and cascade energies. From there, we calculate inelasticity by dividing the reconstructed hadronic energy by the sum of hadronic and track energies. 

\begin{figure}[h]
    \centering
    \includegraphics[width=0.49\linewidth]{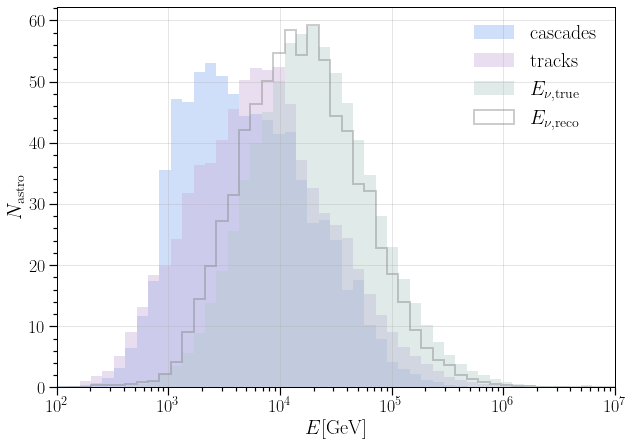}
    \includegraphics[width=0.49\linewidth]{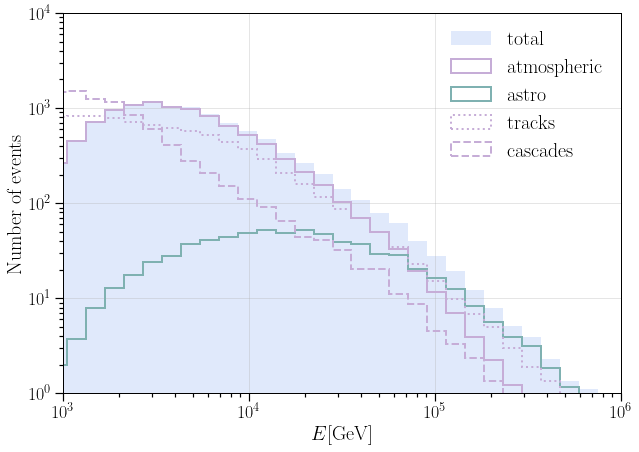}
    \caption{The reconstructed neutrino energy event distributions for different components of the neutrino spectrum. Astrophysical events are shown on the left, and a comparison of astrophysical, atmospheric, and total events is shown on the right.}
    \label{fig:enter-label}
\end{figure}

\subsection{Systematics}

We use a combination of physics and detector systematics to properly model uncertainties. These systematics are interpolated over reconstructed energy, zenith angle, and inelasticity. A summary of the systematics along with priors and allowed ranges is given in Table \ref{tab:systematics}.

The detector systematics that we use to parameterize uncertainties in the IceCube detector include: DOM efficiency, or a scaling factor describing the sensitivity of the individual digital optical modules in the ice \cite{Abbasi_2010}; two hole ice parameters $p_0$ and $p_1$ based on an angular response function \cite{PhysRevD.108.012014} that models effects from ice impurities caused by the columns of refrozen ice surrounding the DOMs \cite{Rongen:2016sbk, fiedlschuster2019effectholeicepropagation}; and scattering and absorption coefficients \cite{Askebjer_1997} used in simulations to account for depth-dependent uncertainties in photon propagation in the bulk ice \cite{Ackermann:2006pva,Aartsen_2013} .

We model the atmospheric neutrino flux uncertainty according to the Global Spline Fit (GSF) parameters for the cosmic ray spectrum as well as the Data-Driven Model (DDM) describing hadronic interaction uncertainties. Together, these are known as the \texttt{daemonflux} parameters \cite{Ya_ez_2023}. The predictions and corresponding uncertainties are very different compared to other fits to data, such as the Barr parameterization \cite{Barr_2006}. The uncertainties are so large in Barr because of an assumed 40\% uncertainty in K$^{\pm}$ production at high energies.  In contrast,  \texttt{daemonflux} relies on a data-driven approach, using data on high-energy (TeV) atmospheric muons to constrain the $\bar{\nu}:\nu$ ratio. There are also slight differences in the choice of primary flux, and differences in kaon abundance can result from different fits. 

Finally, we assume an astrophysical single power law flux as shown here:
\begin{equation}
    \phi^{\rm baseline}_{\rm astro}(E) =\phi_0\Big(\frac{E}{100\;{\rm TeV}}\Big)^{-\gamma},\quad \phi_0 = 1.68\times 10^{-18}\;{\rm GeV}^{-1}{\rm cm}^{-2}{\rm s}^{-1}{\rm sr}^{1},\; \gamma = 2.58.
\end{equation}

\begin{table}[ht!]
\centering
\begin{tabular}{l|c|c|c|l}
Systematic & Central & Prior (1$\sigma$) & Range & Implementation \\
\hline
\multicolumn{5}{l}{Detector Parameters} \\
\hline
DOM efficiency  & 1.27 &$\pm$10\% & [1.234, 1.346] & Spline over 6 knots\\
Hole Ice $p_0$ & -0.3 & $\pm$0.5 &  [-2,1]& Spline over 6 knots \\ 
Hole Ice $p_1$ & -0.1 & $\pm$0.5 &  [-0.2,0.2]& Spline over 6 knots \\ 
Scattering & 1.05 & $\pm$0.05 &  [0.8,1.2]& Spline over 6 knots \\
Absorption & 1.00 & $\pm$0.05 &  [0.8,1.2]& Spline over 6 knots \\ 
\hline
\multicolumn{5}{l}{Conventional Flux Parameters} \\
\hline
K$_{158G}^{\pm}$, $\pi_{20T}^{\pm}$, K$_{2P}^{\pm}$, $\pi_{2P}^{\pm}$, p$_{2P}$, n$_{2P}$ & 0.0 & $\pm$1.0 & [-2, 2] & Spline over 9 knots \\
GSF$_{1-6}$ & 0.0 & $\pm$1.0 & [-4, 4] & Spline over 9 knots
\\
\hline
\multicolumn{5}{l}{High-energy Flux Parameters} \\
\hline
Normalization & 1.68 & $\pm$0.36 & [0,3] & Flux rescaling \\
$\Delta\gamma_{1}$, tilt from -2.5 & 0.0 & $\pm$0.36 & [-2,2] & Flux reweighting \\
\hline
\end{tabular}
\caption{List of systematic parameters considered in this analysis along with their central values, priors, allowed ranges, and implementations. The atmospheric systematics include the Data-Driven Model (DDM) as well as the cosmic-ray Global Spline Fit (GSF) parameters \cite{Ya_ez_2023}.}
\label{tab:systematics}
\end{table}

\begin{figure}[h!]
    \centering
    \includegraphics[width=0.49\linewidth]{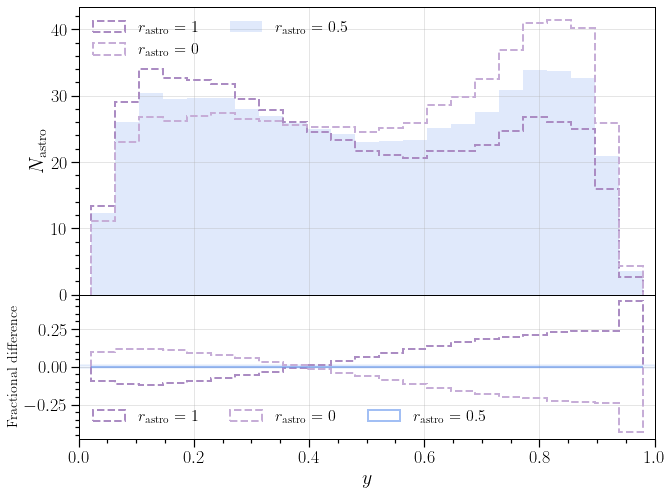}
    \includegraphics[width=0.49\linewidth]{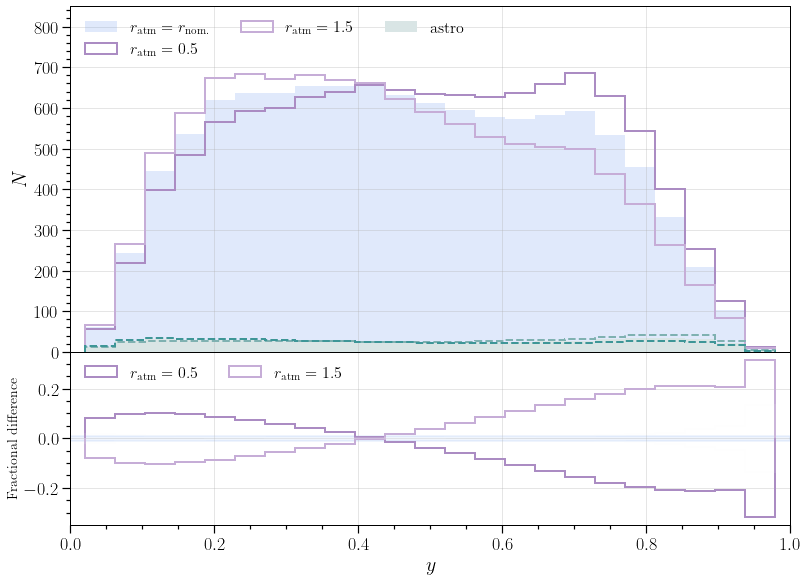}
    \caption{Left: The inelasticity distributions of astrophysical events under different assumptions of $r_{\rm astro}$. The bottom panel shows the fractional difference between each scenario and the nominal scenario, in which $r_{\rm astro} = 0.5$. Right: The total inelasticity distributions under different assumptions of $r_{\rm atm}$. The bottom panel shows the fractional difference between each of these scenarios and the nominal model, in which $r_{\rm atm} = 1$.}
    \label{fig:event-distributions}
\end{figure}

\subsection{Fit Procedure}

For each point in the physics parameter space of $r_{\rm astro} \equiv \big(\nu/(\nu+\bar{\nu})\big)_{\rm astro}$ and $r_{\rm atm} \equiv \Big(\frac{\nu}{\nu+\bar{\nu}}\Big)\Big/\Big(\frac{\nu}{\nu+\bar{\nu}}\Big)_{\rm nominal}$, we generate a predicted event distribution.
We then perform the fit using a binned Poisson likelihood method, where the test statistic is 
\begin{equation}
    -2\ln \Lambda = \sum_{i}\Big(\mu_i(\theta)-n_i+n_i\ln\big(\frac{n_i}{\mu_i(\theta)}\big)\Big)+\sum_{j}\frac{(\theta_j-\theta_j^*)^2}{\sigma_j^2},
\end{equation}
where $\mu_i(\theta)$ is the expected event count per bin and $n_i$ is the observed event count. The second term is a Gaussian penalty for systematic parameter $\theta_j$ with mean $\theta_j^*$ and standard deviation $\sigma_j$. We minimize $-2\ln \Lambda$  over our set of physics and systematic parameters, denoted by $\theta$, using $\texttt{Minuit}$ \cite{James:1975dr}. 
\section{Sensitivity}

\begin{figure}[h!]
    \centering
    \includegraphics[width=0.70\linewidth]{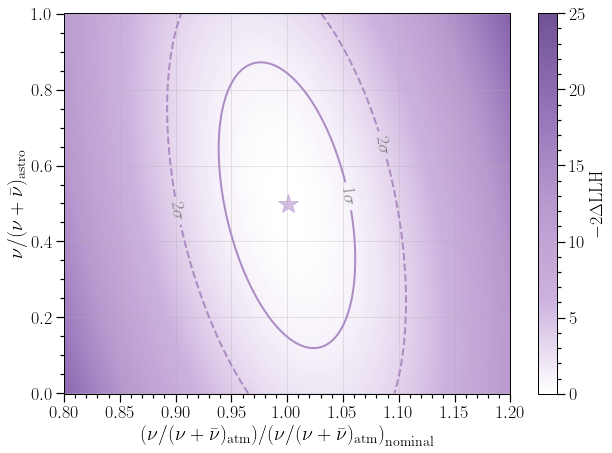}
    \caption{The Asimov sensitivity obtained using our procedure when injecting $r_{\rm astro} = 0.5$ and $r_{\rm atm} = 1$.}
    \label{fig:Results}
\end{figure}

Figure \ref{fig:Results} shows the Asimov sensitivity that we obtain by injecting a nominal model that assumes $r_{\rm astro} = 0.5$ and $r_{\rm atm} = 1$ and performing a fit over a $25\times 25$ grid over $0\leq r_{\rm astro}\leq 1$ and $0.5\leq r_{\rm atm}\leq 1.5$.

\section{Conclusion}

The projected sensitivity shown in Figure \ref{fig:Results} predicts a closed contour at $1\sigma$ with 10.3 years of IceCube data, indicating that IceCube has sensitivity to the $\bar{\nu}:\nu$ ratio through inelasticity measurements. 
Once completed, this produce will result in the first measurement of the astrophysical $\bar{\nu}:\nu$ ratio ever, outside of the single Glashow event detected by IceCube. This result will shed light on numerous questions about the astrophysical neutrino flux, including which production mechanisms that could be involved. 

\bibliographystyle{ICRC}
\bibliography{references}

\providecommand{\href}[2]{#2}\begingroup\raggedright\begin{thebibliography}{10}

\bibitem{Barger:2014iua}
V.~Barger, L.~Fu, J.~G. Learned, D.~Marfatia, S.~Pakvasa, and T.~J. Weiler, \href{http://dx.doi.org/10.1103/PhysRevD.90.121301}{{\em Phys. Rev. D} {\bfseries 90} (2014) 121301}.

\bibitem{IceCube:2021rpz}
{\bfseries IceCube} Collaboration, M.~G. Aartsen {\em et~al.}, \href{http://dx.doi.org/10.1038/s41586-021-03256-1}{{\em Nature} {\bfseries 591} no.~7849, (2021) 220--224}. [Erratum: Nature 592, E11 (2021)].

\bibitem{Klein:2012ug}
S.~R. Klein, R.~E. Mikkelsen, and J.~Becker~Tjus, \href{http://dx.doi.org/10.1088/0004-637X/779/2/106}{{\em Astrophys. J.} {\bfseries 779} (2013) 106}.

\bibitem{Klein:2019nbu}
S.~R. Klein, {\em {Probing high-energy interactions of atmospheric and astrophysical neutrinos}}.
\newblock 2020.
\newblock \href{http://arxiv.org/abs/1906.02221}{{\ttfamily arXiv:1906.02221 [astro-ph.HE]}}.

\bibitem{IceCube:2018pgc}
{\bfseries IceCube} Collaboration, M.~G. Aartsen {\em et~al.}, \href{http://dx.doi.org/10.1103/PhysRevD.99.032004}{{\em Phys. Rev. D} {\bfseries 99} (2019) 032004}.

\bibitem{IceCubeCollaborationSS:2025zgz}
{\bfseries IceCube} Collaboration, R.~Abbasi {\em et~al.}, \href{http://dx.doi.org/10.1103/PhysRevD.111.112001}{{\em Phys. Rev. D} {\bfseries 111} (2025) 112001}.

\bibitem{Formaggio_2012}
J.~A. Formaggio and G.~P. Zeller, \href{http://dx.doi.org/10.1103/revmodphys.84.1307}{{\em Reviews of Modern Physics} {\bfseries 84} (2012) 1307–1341}.

\bibitem{FSR}
R.~Plestid and B.~Zhou, \href{http://dx.doi.org/10.1103/PhysRevD.111.043007}{{\em Phys. Rev. D} {\bfseries 111} no.~4, (2025) 043007}.

\bibitem{ct18}
T.-J. Hou {\em et~al.}, \href{http://dx.doi.org/10.1103/PhysRevD.103.014013}{{\em Phys. Rev. D} {\bfseries 103} (2021) 014013}.

\bibitem{lhapdf}
A.~Buckley, J.~Ferrando, S.~Lloyd, K.~Nordstr{\"o}m, B.~Page, M.~R{\"u}fenacht, M.~Sch{\"o}nherr, and G.~Watt, \href{http://dx.doi.org/10.1140/epjc/s10052-015-3318-8}{{\em Eur. Phys. J. C} {\bfseries 75} (2015) 132}.

\bibitem{Halzen:2010yj}
F.~Halzen and S.~R. Klein, \href{http://dx.doi.org/10.1063/1.3480478}{{\em Rev. Sci. Instrum.} {\bfseries 81} (2010) 081101}.

\bibitem{Aartsen_2017}
{\bfseries IceCube} Collaboration, M.~Aartsen {\em et~al.}, \href{http://dx.doi.org/10.1088/1748-0221/12/03/p03012}{{\em Journal of Instrumentation} {\bfseries 12} no.~03, (2017) P03012–P03012}.

\bibitem{IceCube:2024fxo}
{\bfseries IceCube} Collaboration, R.~Abbasi {\em et~al.}, \href{http://dx.doi.org/10.1103/PhysRevD.110.022001}{{\em Phys. Rev. D} {\bfseries 110} (2024) 022001}.

\bibitem{Aartsen_2014}
{\bfseries IceCube} Collaboration, M.~Aartsen {\em et~al.}, \href{http://dx.doi.org/10.1088/1748-0221/9/03/p03009}{{\em Journal of Instrumentation} {\bfseries 9} no.~03, (2014) P03009–P03009}.

\bibitem{random-forests}
J.~Breiman, \href{http://dx.doi.org/10.1023/A:1010933404324}{{\em Machine Learning} {\bfseries 45} (2001) 5--32}.

\bibitem{Abbasi_2010}
{\bfseries IceCube} Collaboration, R.~Abbasi {\em et~al.}, \href{http://dx.doi.org/10.1016/j.nima.2010.03.102}{{\em Nucl. Instrum. Meth. A} {\bfseries 618} (2010) 139--152}.

\bibitem{PhysRevD.108.012014}
{\bfseries IceCube} Collaboration, R.~Abbasi {\em et~al.}, \href{http://dx.doi.org/10.1103/PhysRevD.108.012014}{{\em Phys. Rev. D} {\bfseries 108} (2023) 012014}.

\bibitem{Rongen:2016sbk}
M.~Rongen, \href{http://dx.doi.org/10.1051/epjconf/201611606011}{{\em EPJ Web Conf.} {\bfseries 116} (2016) 06011}.

\bibitem{fiedlschuster2019effectholeicepropagation}
S.~Fiedlschuster, ``{The Effect of Hole Ice on the Propagation and Detection of Light in IceCube},'' 2019.

\bibitem{Askebjer_1997}
{Askebjer P. {\it et al.}}, \href{http://dx.doi.org/10.1364/ao.36.004168}{{\em Applied Optics} {\bfseries 36} (1997) 4168}.

\bibitem{Ackermann:2006pva}
{\bfseries IceCube} Collaboration, M.~Ackermann {\em et~al.}, \href{http://dx.doi.org/10.1029/2005JD006687}{{\em J. Geophys. Res.} {\bfseries 111} no.~D13, (2006) D13203}.

\bibitem{Aartsen_2013}
{\bfseries IceCube} Collaboration, M.~G. Aartsen {\em et~al.}, \href{http://dx.doi.org/10.1016/j.nima.2013.01.054}{{\em Nucl. Instrum. Meth. A} {\bfseries 711} (2013) 73--89}.

\bibitem{Ya_ez_2023}
J.~P. Ya{\~n}ez and A.~Fedynitch, \href{http://dx.doi.org/10.1103/PhysRevD.107.123037}{{\em Phys. Rev. D} {\bfseries 107} (2023) 123037}.

\bibitem{Barr_2006}
G.~D. Barr, T.~K. Gaisser, S.~Robbins, and T.~Stanev, \href{http://dx.doi.org/10.1103/PhysRevD.74.094009}{{\em Phys. Rev. D} {\bfseries 74} (2006) 094009}.

\bibitem{James:1975dr}
F.~James and M.~Roos, \href{http://dx.doi.org/10.1016/0010-4655(75)90039-9}{{\em Comput. Phys. Commun.} {\bfseries 10} (1975) 343--367}.

\end{thebibliography}\endgroup

%

\clearpage

\section*{Full Author List: IceCube Collaboration}

\scriptsize
\noindent
R. Abbasi$^{16}$,
M. Ackermann$^{63}$,
J. Adams$^{17}$,
S. K. Agarwalla$^{39,\: {\rm a}}$,
J. A. Aguilar$^{10}$,
M. Ahlers$^{21}$,
J.M. Alameddine$^{22}$,
S. Ali$^{35}$,
N. M. Amin$^{43}$,
K. Andeen$^{41}$
% ...
\\
\\
% ...
$^{10}$ Universit{\'e} Libre de Bruxelles, Science Faculty CP230, B-1050 Brussels, Belgium \\
% ...
$^{16}$ Department of Physics, Loyola University Chicago, Chicago, IL 60660, USA \\
$^{17}$ Dept. of Physics and Astronomy, University of Canterbury, Private Bag 4800, Christchurch, New Zealand \\
% ...
$^{21}$ Niels Bohr Institute, University of Copenhagen, DK-2100 Copenhagen, Denmark \\
$^{22}$ Dept. of Physics, TU Dortmund University, D-44221 Dortmund, Germany \\
% ...
$^{35}$ Dept. of Physics and Astronomy, University of Kansas, Lawrence, KS 66045, USA \\
% ...
$^{39}$ Dept. of Physics and Wisconsin IceCube Particle Astrophysics Center, University of Wisconsin{\textemdash}Madison, Madison, WI 53706, USA \\
% ...
$^{41}$ Department of Physics, Marquette University, Milwaukee, WI 53201, USA \\
% ...
$^{43}$ Bartol Research Institute and Dept. of Physics and Astronomy, University of Delaware, Newark, DE 19716, USA \\
% ...
$^{63}$ Deutsches Elektronen-Synchrotron DESY, Platanenallee 6, D-15738 Zeuthen, Germany \\
$^{\rm a}$ also at Institute of Physics, Sachivalaya Marg, Sainik School Post, Bhubaneswar 751005, India \\
% ...

\subsection*{Acknowledgments}

\noindent
The authors gratefully acknowledge the support from the following agencies and institutions:
USA {\textendash} U.S. National Science Foundation-Office of Polar Programs,
U.S. National Science Foundation-Physics Division,
U.S. National Science Foundation-EPSCoR,
U.S. National Science Foundation-Office of Advanced Cyberinfrastructure,
Wisconsin Alumni Research Foundation,
Center for High Throughput Computing (CHTC) at the University of Wisconsin{\textendash}Madison,
Open Science Grid (OSG),
Partnership to Advance Throughput Computing (PATh),
Advanced Cyberinfrastructure Coordination Ecosystem: Services {\&} Support (ACCESS),
Frontera and Ranch computing project at the Texas Advanced Computing Center,
U.S. Department of Energy-National Energy Research Scientific Computing Center,
Particle astrophysics research computing center at the University of Maryland,
Institute for Cyber-Enabled Research at Michigan State University,
Astroparticle physics computational facility at Marquette University,
NVIDIA Corporation,
and Google Cloud Platform;
Belgium {\textendash} Funds for Scientific Research (FRS-FNRS and FWO),
FWO Odysseus and Big Science programmes,
and Belgian Federal Science Policy Office (Belspo);
Germany {\textendash} Bundesministerium f{\"u}r Forschung, Technologie und Raumfahrt (BMFTR),
Deutsche Forschungsgemeinschaft (DFG),
Helmholtz Alliance for Astroparticle Physics (HAP),
Initiative and Networking Fund of the Helmholtz Association,
Deutsches Elektronen Synchrotron (DESY),
and High Performance Computing cluster of the RWTH Aachen;
Sweden {\textendash} Swedish Research Council,
Swedish Polar Research Secretariat,
Swedish National Infrastructure for Computing (SNIC),
and Knut and Alice Wallenberg Foundation;
European Union {\textendash} EGI Advanced Computing for research;
Australia {\textendash} Australian Research Council;
Canada {\textendash} Natural Sciences and Engineering Research Council of Canada,
Calcul Qu{\'e}bec, Compute Ontario, Canada Foundation for Innovation, WestGrid, and Digital Research Alliance of Canada;
Denmark {\textendash} Villum Fonden, Carlsberg Foundation, and European Commission;
New Zealand {\textendash} Marsden Fund;
Japan {\textendash} Japan Society for Promotion of Science (JSPS)
and Institute for Global Prominent Research (IGPR) of Chiba University;
Korea {\textendash} National Research Foundation of Korea (NRF);
Switzerland {\textendash} Swiss National Science Foundation (SNSF).

\end{document}